\documentclass{aa} 

\usepackage{graphicx}
\usepackage{color}

\begin{document}

\title{Planet formation from the ejecta of common envelopes}
\titlerunning{Planet formation from the ejecta of common envelopes} 
\authorrunning{Schleicher \& Dreizler}

\author
  {Dominik R.\,G. Schleicher
  \inst{1}
  \and
  Stefan Dreizler
  \inst{1}
  }

\institute{Institut f\"ur Astrophysik, Georg-August-Universit\"at G\"ottingen, Friedrich-Hund-Platz 1, D-37077 G\"ottingen, Germany \\
\email{dschleic@astro.physik.uni-goettingen.de, dreizler@astro.physik.uni-goettingen.de}
}

\date{\today}


\abstract
{The close binary system NN Serpentis must have gone through a common envelope phase before the formation of its white dwarf. During this phase, a substantial amount of mass was lost from the envelope. The recently detected orbits of circumbinary planets are likely inconsistent with planet formation before the mass loss.}
{We explore whether new planets may have formed from the ejecta of the common envelope and derive the expected planetary mass as a function of radius.}
{We employed the Kashi \& Soker model to estimate the amount of mass that is retained during the ejection event and inferred the properties of the resulting disk from the conservation of mass and angular momentum. The resulting planetary masses were estimated from models with and without radiative feedback. }
{We show that the observed planetary masses can be reproduced for appropriate model parameters. {Photoheating can stabilize the disks in the interior, potentially explaining the observed planetary orbits on scales of a few AU.} We compare the expected mass scale of planets for 11 additional systems with observational results and find hints of two populations, one consistent with planet formation from the ejecta of common envelopes and the other a separate population that may have formed earlier. }
{The formation of the observed planets from the ejecta of common envelopes seems feasible. The model proposed here can be tested through refined observations of additional post-common envelope systems. {While it appears observationally challenging to distinguish between the accretion on pre-existing planets and their growth from new fragments, it may be possible to further constrain the properties of the protoplanetary disk through additional observations of current planetary candidates and post-common envelope binary systems. }}

\maketitle

\section{Introduction}
Within the past few years, an increasing number of planets have been discovered orbiting both unevolved and post-common envelope binaries (PCEBs), and many additional systems are currently suspected of hosting planets. For instance, the unevolved dG/dM binary Kepler~47 harbors two planets orbiting the system with semi-major axes of less than $1$~AU \citep{Orosz12}. The binary star system Kepler~16 hosts the Saturn-sized planet Kepler~16b with a semi-major axis of $3.9$~AU \citep{Doyle11}, and Kepler~34/35 hosts a planet with one fifth of Jupiter's mass at a semi-major axis of $1$~AU on a somewhat eccentric orbit \citep{Welsh12}. For the PCEB system NN Ser, \citet{Beuermann10, Beuermann13} employed the light-travel time (LLT) effect to detect two planets with masses of $7.0\ M_J$ and $1.7\ M_J$ with a semi-major axis of $5.4$~AU and $3.47$~AU, respectively. They also demonstrate the dynamical stability of these orbits, which was independently confirmed by \citet{Horner12} {and \citet{Marsh13}}. Recent data by \citet{Parsons13} further rule out apsidal precession as an alternative interpretation, and further strenghten the previous conclusions.

 In the system HW~Vir, two planets have passed the test of secular stability \citep{Beuermann12}. A final conclusion about the stability of planetary orbits in Hu~Aqr is currently pending \citep{Godziewski12, Hinse12}, and more data are required to understand the eclipse-time variations in QS~Vir \citep{Parsons10b}. Overall, \citet{Zorotovic13} list 12 planet candidates in PCEB systems. {A set of planet formation scenarios in highly evolved systems, including binaries, has recently been discussed by \citet{Perets10} and \citet{Tutukov12}.}

The recent detections of PCEB planets have raised interesting questions concerning the origin of the observed planets. From a theoretical perspective, they may have formed either before the common envelope (CE) event and survived the subsequent mass loss, or they may have formed from material ejected during the CE phase. For NN Ser, the first of these possibilities has been explored by \citet{Volschow13}, showing that the observed planetary orbits are likely inconsistent with this scenario. {For this purpose, they followed the evolution of planetary orbits by assuming a rapid ejection event and find that the planets should likely be ejected or arrive at highly eccentric orbits, which are not observed. In case the initial orbits are tuned to avoid the eccentricity problem, they turn out to be dynamically unstable. Their calculation neglects the impact of friction on the orbit, which they estimated by assuming spherical and disk-like geometries for the outflowing gas. In a similar manner, \citet{Mustill13} show that the main-sequence progenitor configurations of NN Ser would be dynamically unstable, implying that the planets would not have survived the main sequence phase. These results suggest that PCEB planets manifest a second phase of planet formation.} As the hot white dwarf in NN Ser has a cooling age of only $10^6$~yrs \citep{Parsons10a}, the planets can be expected to be dynamically young and may be the youngest planets detected so far. {As a caveat, we mention however  that the analysis by \citet{Volschow13} and \citet{Mustill13} did not account for mass and angular momentum accretion of the planets after the ejection phase. There is thus a valid possibility that the previous planets became the cores of the new planets formed from the ejecta of the envelope.} {The young age of the system also makes it unlikely that the planets formed from the coagulation of planetesimals, which typically occurs on timescales of $10-100$~Myrs \citep{Pollack96, Weidenschilling00, Kenyon14}. Due to the close-to-spherical orbits, the system NN Ser provides a particularly strong case against a purely orbital change of pre-existing planets, while an eccentricity of $20\%$ as in HU Aqr is potentially still compatible with semi-adiabatic orbital expansion \citep{Zwart13}.}

The CE model originally put forward by \citet{Paczynski76} is the central ingredient for the formation of many close binary systems, as discussed for instance by \citet{Meyer79}, \citet{Iben93}, \citet{Taam00}, \citet{Webbink08} and \citet{Taam10}. A CE is defined as a structure where two stars share the envelope. One of them is typically a giant, for instance in an AGB phase \citep{Herwig05}, while the companion can be a white dwarf or a main-sequence star. In this phase, the orbital period decreases due to gravitational drag and tidal interactions \citep[e.g.,][]{Iben93, Kashi11}, while both the energy and angular momentum of the companion are injected into the envelope. {A recent review of the CE evolution was provided by \citet{Ivanova13}.}

There is, however, still a controversy over the timescales of these processes. For instance, \citet{Livio88} and \citet{Rasio96} suggest that the energy is deposited in a very short time, because most of the gravitational energy is released at small orbital separations where the radius decreases quickly. On the other hand, \citet{Sandquist98} estimate that the overall timescale of the process  lasts $\sim200$~days, while \citet{DeMarco03, DeMarco09} derived timescales of $9-18$~yrs until a negligible amount of material remained in the envelope. The numerical simulations by \citet{Passy12}, on the other hand, indicate a typical timescale of $100-200$~days. {Since the gravitational potential energy of the companion star scales as $a^{-1}$, where $a$ denotes its distance from the core of the red giant, the energy deposition rate in a given range $da$ scales as $a^{-2}$. We therefore expect that the most efficient energy deposition occurs on scales close to the core, where the timescales are shortest, while hydrodynamical simulations may find that the complete process takes longer, because some of the energy is deposited at earlier stages. Since a large portion of the energy is thus deposited in a short time close to the central core, we adopt the instantaneous injection approximation developed by \citet{Kashi11} in this manuscript. A consideration of the enthalpy indeed shows that the material close to the core can be evacuated on short timescales, thus favoring the survival of low-mass companions by preventing a merger with the core \citep{Ivanova11}. }

Owing to the deposition of  energy and angular momentum in the envelope, a substantial fraction, or almost the entire envelope, can be removed from the star. An important controversy, however, concerns the question of how much of this material remains gravitationally bound to the star. A first calculation by \citet{Sandquist98} indicated  that in the case of a $5$~$M_\odot$ AGB star interacting with a $0.6$~$M_\odot$ companion, the companion unbinds $1.55$~$M_\odot$ or $23\%$ of the AGB envelope. Similarly, \citet{Passy12} find in numerical simulations that $80\%$ of the envelope mass remains bound to the star. The latter may correspond to an upper limit, since their calculation did not consider the potential rotation of the envelope. An analytical assessment by \citet{Kashi11}, on the other hand, indicates that $1-10\%$ of the envelope mass is retained. {Their model yields a bound fraction of $5\%$ for their fiducial case, a red giant with $5$~M$_\odot$, and a larger fraction of $\sim10\%$ for a system like NN Ser. The last  provides a lower limit on the total mass that is retained, because the typical velocities may be reduced in the case of a more gradual injection. In addition, deviations from spherical symmetry would imply that part of the gas carries away larger fractions of the kinetic energy, therefore increasing the gas fraction that remains bound to the star.}

The implications of rotation {during the common envelope phase} have been explored  by \citet{Sandquist98}, who showed that a differentially rotating structure, similar to a thick disk, surrounds the binary during an intermediate phase of the CE evolution. A similar thick disk structure was analytically obtained by \citet{Soker92, Soker04} and \citet{DeMarco11}.  While both {\citet{Soker92, Soker04} and \citet{Sandquist98}}  considered the envelope before the ejection, \citet{DeMarco11} propose that the part of the ejected envelope that remains gravitationally bound to the system may fall back and form a circumbinary disk. The implications of such a disk for the final fate of the star and for type~Ia~supernovae have been explored by \citet{Kashi11}.

In this study, we go one step further and consider the implications of self-gravitating instabilities in these disks for the formation of a new generation of planets. In section~\ref{ejection}, we employ the model of \citet{Kashi11} for the ejection of mass in the common envelope phase and calculate which fraction of the gas remains gravitationally bound to the system. We further estimate the total angular momentum that is available for the formation of a disk. In section~\ref{grav}, we assume that such a disk has formed and estimate the expected planetary masses resulting from gravitational instabilities. The implications of radiative feedback from the central star are considered in section~\ref{rad}. Potential implications for other systems are explored in section~\ref{other}, and our results are summarized and discussed in section~\ref{discussion}.

\section{Ejection of mass and angular momentum}\label{ejection}
We consider a binary system consisting of an AGB star and a low-mass companion with masses $M_1$ and $M_2$ and orbital separation $a$. As our fiducial system, we adopt NN~Ser, which currently consists of a white dwarf of $0.535$~M$_\odot$ and a companion star with $M_2=0.111$~M$_\odot$. {Using the binary star evolution code developed by \citet{Hurley02}, \citet{Mustill13}  show that the expected mass of the progenitor ranges between $1.875$~M$_\odot$ and $2.25$~M$_\odot$ for metallicities of $Z=0.01-0.03$, thus indicating a mass loss factor $\mu=M_{current}/M_{prev}\sim0.23-0.29$. In the following, we  assume a generic progenitor mass of $M_1=2$~M$_\odot$ for NN Ser. Adopting a core mass of $M_{core}=0.535$~M$_\odot$, which corresponds to the mass of the white dwarf, the ejection of the envelope follows from the model of \citet{Kashi11} as shown below. For the AGB star, we adopt the envelope structure from the models of \citet{Kashi11}, \citet{Nordhaus06}, \citet{Tauris01} and \citet{Soker92} by assuming a power-law profile} \begin{equation}
\rho(r)=Ar^{-\omega}
\end{equation}
with $\omega=2$. The remaining mass $M_{core}$ is assumed to be centrally concentrated in a region smaller than $0.01$~$R_\odot$. The enclosed mass within radius $r$ is thus\begin{equation}
M(r)=M_{core}+\int_{R_{core}\sim0}^r 4\pi r^2\rho(r) dr=M_{core}+4\pi A r.
\end{equation}
The normalization of the density profile is then\begin{equation}
A=\frac{M_1-M_{core}}{4\pi R_*},
\end{equation}
where $R_*$ denotes the radius of the AGB star. The latter is estimated from the model of \citet{Kashi11}, assuming a scaling relation with $M_1^{0.5}$, yielding $R_*\sim185R_\odot$. 

\begin{figure}[htbp]
\begin{center}
\includegraphics[scale=0.5]{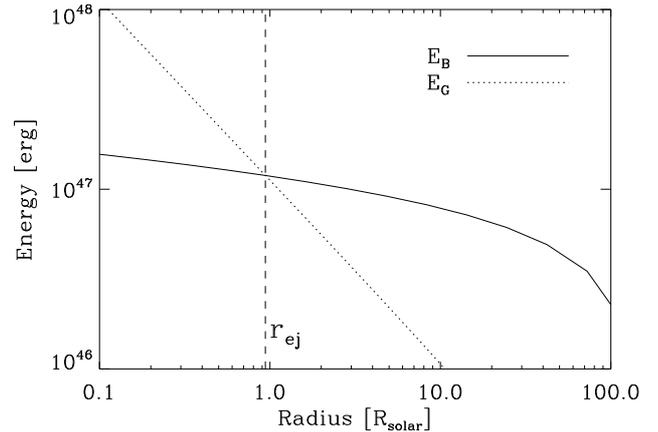}
\caption{Gravitational binding energy of the envelope vs gravitational energy released via the inspiral in our fiducial system. Both energies become equal for $E_B=E_G\sim1.2\times10^{47}$~erg at a radius of $\sim0.94$~$R_\odot$. }
\label{fig:energy}
\end{center}
\end{figure}

When the companion star spirals into the envelope of the giant, we  assume that mass ejection occurs when the integrated binding energy of the envelope mass outside the orbital radius becomes lower than the gravitational energy released during the inspiral. For the model sketched here, the binding energy of the envelope mass outside radius $r$ is given as\begin{eqnarray}
E_B&=&\int_r^{R_*} \frac{G(M(r)+M_2)}{r}4\pi r^2\rho(r)dr\\
&=&4\pi AG(M_{core}+M_2)\ln\left(\frac{R_*}{r}\right)+16\pi^2 GA^2(R_*-r).\nonumber
\end{eqnarray}
When the companion star spirals from the surface of the AGB down to a separation $a$, the released gravitational energy is\begin{eqnarray}
E_G&=&\frac{GM(a)M_2}{2a}-\frac{GM_1M_2}{2R_*}\\
&=&\frac{GM_2 M_{core}}{2a}+2\pi G M_2 A-\frac{GM_1 M_2}{2R_*}.\nonumber
\end{eqnarray}
The factor $1/2$ in this expression reflects that the total energy of the system corresponds to half of the gravitational potential energy (virial theorem). We note that the envelope does not effectively contribute to the release of the gravitational energy, since the envelope mass scales linearly with radius for the adopted profiles, implying an equal contribution in the initial and the final stages. For our fiducial system, the gravitational and binding energies become equal at a separation $r_{ej}\sim0.9$~$R_\odot$, in agreement with the observed value $a_{NN}=0.934$~$R_\odot$ \citep{Parsons10a}. The corresponding released binding energy is $E_B=E_G\sim1.2\times10^{47}$~erg. The  evolution of the energies as a function of radius is given in Fig.~\ref{fig:energy}. As the envelope outside $r_{ej}$ becomes unbound, the ejected mass is $M_{ej}\sim 4\pi A(R_*-r_{ej})\sim1.24$~$M_\odot$.

Central questions are, however, which fraction of the released mass becomes truly unbound and therefore escapes from the binary system and which fraction can be retained and subsequently contribute to the formation of a new generation of planets. To estimate the fraction of the mass that is retained, we adopt the approach of \citet{Kashi11}, who assumed that the released gravitational energy is instantaneously ejected at the radius $r_{ej}$. Under the additional assumption of spherical symmetry, the propagation of the resulting blast wave can be followed by employing the self-similar solution of \citet{Sedov59}. While the assumption of spherical symmetry is certainly an approximation, the calculation by \citet{Kashi11} indeed confirmed that the timescale for the blast wave to reach the surface of the common envelope is considerably longer than the time for the inspiral on the scale $r_{ej}$, implying that the assumption of instantaneous injection is reasonable. When assuming the injection of energy $E_0$ at $t=0$ in the center of a density profile $\rho(r)=Ar^{-\omega}$, the position of the shock front is then \begin{equation}
R_S(t)=\left(\frac{E_0t^2}{\alpha A}  \right)^{1/(5-\omega)}.
\end{equation}
As noted by \citet{DeMarco11} and \citet{Kashi11}, both the released gravitational energy and the thermal energy of the envelope contribute to the expansion. The virial theorem means that we thus have $E_{th}=0.5 E_B$ and $E_0=E_B+E_{th}=\frac{3}{2}E_B$. The constant $\alpha$ in the solution follows from energy conservation, as we show below. The time $t_f$ when the shock front reaches the radius $R_*$ of the giant follows  from  $R_S(t_f)=R_*$, yielding\begin{equation}
t_f=\sqrt{\frac{\alpha A}{E_0}}R_*^{(5-\omega)/2}.
\end{equation}
The propagation velocity of the shock follows as\begin{equation}
V_S=\dot{R}_S=\frac{2}{5-\omega}\left( \frac{E_0}{\alpha A} \right)^{1/(5-\omega)}t^{(\omega-3)/(5-\omega)}.
\end{equation} 
The density, velocity and pressure behind the shock are then $v=v_S \lambda$, $\rho=\rho_S \lambda$, and $p=p_S\lambda$, where $v_S$, $\rho_S$, and $p_S$ denote the quantities immediately behind the shock front, and the self-similar variable $\lambda$ is \begin{equation}
\lambda=\left(  \frac{A\alpha}{E_0}\right)^{1/(5-\omega)}rt^{-2/(5-\omega)}=r/R_S.
\end{equation} 
For blast waves in a density profile $\rho(r)=Ar^{-\omega}$ and an adiabatic index $\gamma$, the analytic solution is particularly simple in the case where $\omega=(7-\gamma)/(\gamma+1)$. This is true in particular for our model of the AGB star with $\omega=2$ and $\gamma=5/3$. Combined with the strong-shock jump conditions \begin{eqnarray}
\rho_S(t)&=&\frac{\gamma+1}{\gamma-1}\rho(R_S(t)),\\
v_S(t)&=&\frac{2}{\gamma+1}V_s(t),\\
p_S(t)&=&\frac{2}{\gamma+1}\rho(R_S(t))V_s^2(t),
\end{eqnarray}
we obtain the post-shock profiles 
\begin{eqnarray}
v&=&\frac{2}{3\gamma-1}\frac{r}{t}=\frac{1}{2}\frac{r}{t},\\
\rho&=&\frac{A(\gamma+1)}{r^\omega(\gamma-1)}\lambda^{8/(\gamma+1)}=4Ar^{-2}\lambda^3,\\
p&=&\frac{A}{r^{\omega-2}t^2}\frac{2(\gamma+1)}{(3\gamma-1)^2}\lambda^{8/(\gamma+1)}=\frac{1}{3}At^{-2}\lambda^{3},
\end{eqnarray}
where we employed $\omega=2$ and $\gamma=5/3$ after the second equality. {We note that the solution derived here becomes invalid after breakout from the star, because the assumed density profile only holds in the interior.} When the shock reaches the stellar radius $R_*$, the total kinetic energy involved in the ejection is\begin{equation}
E_{kin}=\int_{r_{ej}\sim0}^{R_*} 4\pi r^2 \rho(r,t_f)v^2(r,t_f)dr=\frac{2\pi}{3}\frac{E_0}{\alpha}.\end{equation}
Considering that the internal energy density is related to the pressure via $e=\frac{3}{2}p$, the thermal energy in the expanding envelope is\begin{equation}
E_{therm}=\int 4\pi r^2\frac{3}{2}p(r,t_f)dr=\frac{\pi}{3}\frac{E_0}{\alpha}.
\end{equation}
Considering that $E_{therm}+E_{kin}=E_0$, the normalization follows as $\alpha=\pi$. To determine the fraction of the mass that remains gravitationally bound to the binary system, we compare the velocity profile at $t=t_f$, \begin{equation}
v(r,t_f)=\frac{1}{2}r\left( \frac{\alpha AR_*^3}{E_0} \right)^{-1/2},
\end{equation}
with the escape velocity of the system. Because $r_c\ll R_*$, the remaining mass of the AGB star is essentially the mass of the core. Since  the secondary also contributes through its gravitational mass, the escape velocity is\begin{equation}
v_{esc}=\left( \frac{2G(M_c+M_2)}{r} \right)^{1/2}.
\end{equation}
From this comparison, we determine that the gas below a scale of \begin{equation}
r_b=2^{2/3}\left( 2G(M_{core}+M_2) \right)^{2/3}\left( \frac{\alpha AR_*^3}{E_0} \right)^{1/3}
\end{equation}
remains gravitationally bound. This mass is then\begin{equation}
M_{bound}=\int_0^{r_b} 4\pi r^2 \rho(r,t_f)dr=4\pi A\frac{r_b^4}{R_*^3}.
\end{equation}
For our fiducial system NN Ser, we find $r_b\sim106$~$R_\odot$ and $M_{bound}\sim0.133\ M_\odot$, corresponding to $140$ Jupiter masses. As this derivation was, however, based on simplifying assumptions, we parametrize the total mass available for disk formation as \begin{equation}
M_{disk}=\alpha_M M_{bound},\label{Mdisk}
\end{equation}
where $\alpha_M$ describes potential corrections arising in more realistic scenarios.

During the inspiral, the progenitor star deposits its angular momentum in the common envelope. The total angular momentum that is deposited into the envelope can thus be estimated as\begin{equation}
L_{dep}=M_2R_*\sqrt{\frac{GM_1}{R_*}},
\end{equation}
where we adopted the Kepler velocity for the stellar companion. In our fiducial system, the injected angular momentum thus corresponds to $1.2\times10^{52}$~erg~cm$^2$~s$^{-1}$. However, additional angular momentum may be present if the common envelope was previously rotating. {We note that angular momentum can be transported from the companion star into the envelope of the giant via tidal torques before the common envelope phase \citep[see e.g.][]{Hut81, Soker95, Hurley02, Zahn08}. As shown by \citet{Bear10}, the envelope in a system like NN Ser could reach $f_{rot}=45\%$ of its breakup velocity already before the CE. The angular momentum due to the rotation of the envelope is then}\begin{eqnarray}
L_{rot}&=&\int_0^{R_*} 0.45\cdot4\pi r^2 dr \rho(r)r\left( \frac{GM(r)}{r} \right)^{1/2}\nonumber\\
&\sim& 0.9\pi AR_*^2\left( 4\pi AG\right)^{1/2},
\end{eqnarray}
where we assumed that the dominant contribution results from the outer scales of the envelope. For our fiducial system, NN Serpentis, we then have $L_{rot}\sim2.6\times10^{52}$~g~cm$^2$~s$^{-1}$. {The total angular momentum available during the common envelope phase may therefore exceed the angular momentum of the companion by a factor of $3$. For our subsequent considerations, the quantity of interest is, however, the specific angular momentum (i.e., the angular momentum per unit mass) in the material that forms the disk. The latter is parametrized as}\begin{equation}
\frac{L_{disk}}{M_{disk}}=\alpha_L  \frac{L_{dep}}{M_{ej}},\label{Ldisk}
\end{equation}
{where $\alpha_L=f_{rot}f_{inh}$ describes the enhancement of the specific angular momentum in the disk compared to the total system. In this respect, the factor $f_{rot}$ describes the contribution from the rotation of the envelope, which can be of order $3$, while $f_{inh}$ parametrizes the additional enhancement due to the inhomogeneous injection of the angular momentum. Assuming Keplerian rotation of the companion, one can show that its angular momentum scales as $r^{1/2}$, implying an angular momentum deposition $dL/dr\propto r^{-1/2}$ or $dL/dM\propto r^{-1/2}$, as $dM=Adr$. On scales of a few solar radii, we therefore expect an enhancement of the deposited specific angular momentum by a factor of $\sim10$. While the latter shows that a significant enhancement of the specific angular momentum is possible, the precise determination of these parameters is clearly a subject for future investigations. We note, however, that an enhancement of the specific angular momentum by a factor of $10$ was found for the extended gas disk observed in the Red Rectangle \citep{Bujarrabal03, Bujarrabal05}, which was proposed to form from outflows of the post-AGB binary system \citep{Akashi08}.}

\section{Planet formation via gravitational instabilities}\label{grav}

{In the following, we assume that the retained mass $M_{disk}$ settles into a disk and that the distribution of the gas surface density $\Sigma(r)$ follows a power-law profile}
\begin{equation}
\Sigma(r)=\Sigma_0\left( \frac{r_{out}}{r} \right)^n,
\end{equation}
where $r_{out}$ denotes the outer radius of the disk, and $\Sigma_0$ the surface density of the outer radius. {The power-law index $n$ describes the steepness of the profile, where $n=1$ corresponds to a Mestel disk \citep{Mestel63}. The latter will be considered as our fiducial scenario, but we  also explore the effect of different power laws below.} As shown by \citet{Artymowics91}, the innermost stable orbit is typically given as $2.5$ times the binary separation. 

We assume here that the latter already corresponds to its present value of $0.00398$~AU, since the timescales for inspiral are considerably shorter than the timescale for the blastwave to break out \citep{Kashi11}.  However, even for higher values, the contribution of small scales to mass and angular momentum can be neglected, so that we can integrate to $r=0$ for simplicity. Assuming the latter is given, the surface density can be integrated to obtain the total mass of the disk. The normalization of the surface density then follows as\begin{equation}
\Sigma_0=\frac{2-n}{2\pi}\frac{M_{disk}}{ r_{out}^2}.\label{Sigma0}
\end{equation}
To calculate the angular momentum of the disk, we assume that it rotates at its Kepler velocity around a central source of mass $M_{core}+M_2$. One can then show that\begin{eqnarray}
L_{disk}&=&\int_0^{r_{out}}2\pi r dr r \Sigma(r) \sqrt{\frac{G(M_{core}+M_2)}{r}}\nonumber\\
&=&\frac{2-n}{5/2-n}M_{disk}r_{out}^{1/2}\sqrt{G(M_{core}+M_2)},
\end{eqnarray}
where we used Eq.~(\ref{Sigma0}) after the second equality. With $L_{disk}$ determined via Eq.~(\ref{Ldisk}), the outer radius is then\begin{equation}
r_{out}=\left( \frac{5/2-n}{2-n} \right)^2\frac{L_{disk}^2}{M_{disk}^2G(M_{core}+M_2)}.
\end{equation}
Obtaining planets on a scale of $\sim5.4$~AU, as observed in NN~Ser \citep{Beuermann13},  requires that the {specific} angular momentum of the disk  is enhanced by a factor $\alpha_L\sim10$ compared to the {specific angular momentum of the total system (neglecting envelope rotation). As discussed in the previous section, the latter seems feasible due both to the highly inhomogeneous injection of the angular momentum, and to the spin-up of the envelope before the common-envelope phase \citep{Bear10}.}

\begin{figure}[htbp]
\begin{center}
\includegraphics[scale=0.5]{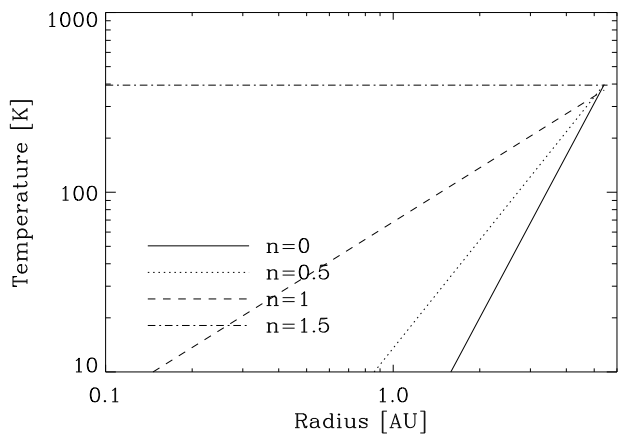}
\caption{{The gas temperature (determined from the condition $Q=1$) as a function of radius for the models defined in Table~\ref{Models}, with power-law slopes from $n=0$ to $n=1.5$.  }}
\label{fig:temp}
\end{center}
\end{figure}

{In this section, we consider disk fragmentation in the absence of additional heat sources like stellar radiation, following the approach of \citet{Levin07}. While the disk cools, the Toomre Q parameter, given as} \begin{equation}
Q=\frac{c_s \Omega}{\pi G\Sigma}\sim1,
\end{equation}
{with $c_s$  the sound speed and $\Omega$ the angular velocity, will decrease. Spiral structures start appearing at $Q\sim1.7$, while fragmentation occurs for $Q\sim1$ \citep{Durisen07, Helled13}.} When this stage is reached, turbulence and shocks are expected to develop, providing a heating mechanism that may compensate for the cooling of the disk \citep{Gammie01}. We can therefore estimate the sound speed at the stage of fragmentation via\begin{equation}
c_s=\frac{\pi G\Sigma}{\Omega}
\end{equation}
and the gas temperature in the midplane of the disk,\begin{equation}
T\sim\frac{2m_p c_s^2}{k_B}=\frac{2m_p}{k_B}\left( \frac{\pi G\Sigma}{\Omega} \right)^2,
\end{equation}
where $m_p$ is the proton mass and $k_B$ Boltzmann's constant (see Fig.~\ref{fig:temp}). Assuming a marginally unstable disk, we can neglect the impact of self-gravity on the disk height, implying that $h(r)=c_s(r)/\Omega(r)$ \citep{Levin07, Lodato07}. We approximate the angular velocity $\Omega(r)$ based on the Kepler law. The {fragmentation mass describes the mass scale of the first clumps forming via the gravitational instability and can be estimated as \citep[see][]{Boley10, Meru10, Meru11, Rogers12}}
\begin{equation}
M_{cl}=\Sigma(r)h^2(r).
\end{equation}
These clumps may, however, substantially grow at a rate of $\dot{M}_{cl}\sim\Omega M_{cl}$ during the dynamical time of the system. The upper limit of the mass that can be reached  is obtained when a gap in the gas disk has formed \citep{Lin86}, yielding a final mass scale of
 \begin{equation}
M_f=M_{cl}\left[ 12\pi\left( \frac{\alpha_{crit}}{0.3} \right) \right]^{1/2}\left(\frac{r}{h}  \right)^{1/2}.
\end{equation}
Here, $\alpha_{crit}$ denotes the critical value of the $\alpha$-parameter for viscous dissipation due to self-gravity at which fragmentation occurs; here, we adopt a generic value of $\alpha_{crit}=0.3$ \citep[e.g.,][]{Gammie01}. {We note  that such clumps will not necessarily grow to their respective isolation masses. In particular, they may rapidly migrate inward \citep{Baruteau11, Michael11, Zhu12}, which can lead to a complete destruction due to tides \citep{Boley10, Nayakshin10} or to changing boundary conditions \citep{Vazan12}. In addition, growth may be delayed by the formation of a circum-protoplanetary disk \citep{Ayliffe12, Helled13}. While a detailed treatment of such effects is beyond the scope of this work, we assume here that the inward migration stops if the planetary masses are higher than the gap-opening mass $M_{gap}$, implying that the orbit of the planet can be cleared from gas, thus strongly reducing the effect of dynamical friction. For the latter, we adopt the expression given by \citet{DAngelo11}, replacing the stellar masses with the total mass $M_{tot}$ of the binary:}
\begin{equation}
\left( \frac{M_{gap}}{M_{tot}} \right)^2=3\pi\alpha f\left( \frac{h}{r} \right)^2\left(\frac{R_H}{r}  \right)^3,
\end{equation}
{where $\alpha\sim\alpha_{crit}$ is a parameter related to the effective disk viscosity, $f$  a parameter of order $1$, $h=c_s/\Omega$ is the disk height, and $R_H$ denotes the Hill radius, given as}
\begin{equation}
R_H=r\left( \frac{M_P}{3M_{tot}} \right)^{1/3}.
\end{equation}
{To see if the planets survive, we set in the following $M_P\sim M_{f}$. While we are aware that the initial fragments are smaller, the study of \citet{Baruteau11} has shown that typical migration times correspond to about ten orbital periods, implying significant growth by a factor of $\sim e^{10}$. As we  see below, this is more than sufficient to reach the final mass $M_f$.}

\begin{figure}[htbp]
\begin{center}
\includegraphics[scale=0.5]{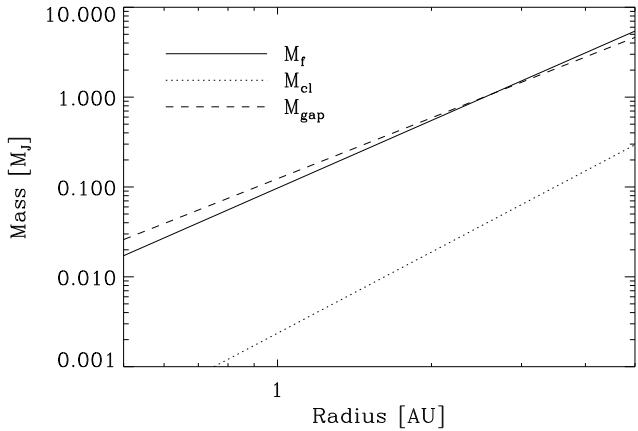}
\caption{The initial and final clump masses, denoted as $M_{cl}$ and $M_f$, as well as the gap opening mass $M_{gap}$ in our fiducial system NN Ser with model parameters $\alpha_M=1.1$, $\alpha_L=12.5$, and $n=1$ in the absence of radiative feedback. The final clump masses are close to the gap opening mass scale, implying that they can evacuate the orbit and therefore ensure their survival.}
\label{fig:clump}
\end{center}
\end{figure}

{For our fiducial system NN Ser, with $\alpha_M\sim1.1$ and $n=1$, we obtain a final mass of $6.6$~$M_J$ on a scale of $5.4$~AU and $1.9\ M_J$ at $3.4$~AU, which is close to the observed values of $7.0\ M_J$ and $1.7\ M_J$ \citep{Beuermann13}. The initial clump masses correspond to $0.37\ M_J$ and $\sim0.09\ M_J$, respectively, at gas temperatures of $370$~K and $231$~K. For a parameter $f=0.5$, the gap-opening masses are $5.5$~M$_J$ and $1.9$~M$_J$, respectively, implying the survivial of the planets. The different mass scales as a function of radius are given in Fig.~\ref{fig:clump}. We note, however, that the parameter $f$ is currently not well constrained and that more detailed numerical investigations will be necessary in the future.}

\begin{table}[htdp]
\begin{center}
\begin{tabular}{c|c|c|c|c}
Model & $n$ & $\alpha_M$ & $\alpha_L$ & $M_{p2}$~[M$_J$]\\
\hline
A & $0$ & $0.57$ & $15$ & $0.7$\\
B & $0.5$ & $0.75$ & $14$ & $1.2$\\
C & $1$ & $1.1$ & $12.5$ & $1.9$\\
D & $1.5$ & $2.3$ & $9.5$ & $3.6$
\end{tabular}
\end{center}
\caption{{Model parameters leading to the formation of a $7.0$~M$_J$ planet at $5.4$~AU in NN Ser, and expected mass $M_{p2}$ for the second planet at $3.47$~AU.}}\label{Models}
\end{table}%

\begin{figure}[htbp]
\begin{center}
\includegraphics[scale=0.5]{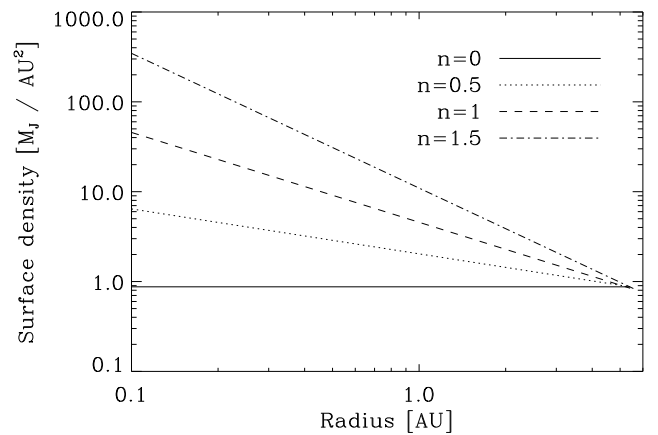}
\caption{{The gas surface density {in Jupiter masses per AU$^2$} as a function of radius for the models defined in Table~\ref{Models}, with power-law slopes from $n=0$ to $n=1.5$.The models are normalized to reproduce the planetary mass at $5.4$~AU, implying that the gas density at this radius is the same.}}
\label{fig:sigma}
\end{center}
\end{figure}

\begin{figure}[htbp]
\begin{center}
\includegraphics[scale=0.5]{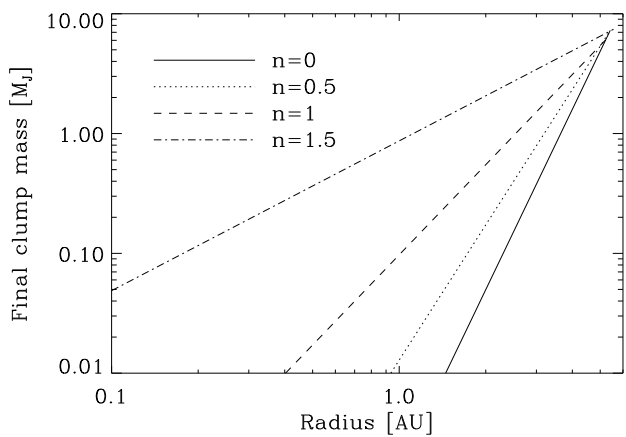}
\caption{{The final clump masses as a function of radius in the absence of radiative feedback for the models defined in Table~\ref{Models}, with power-law slopes from $n=0$ to $n=1.5$. Steeper slopes imply a larger gas reservoir in the interior and thus the formation of more massive clumps. }}
\label{fig:clumpf}
\end{center}
\end{figure}

{To investigate the dependence on the power-law slope $n$ of the gas surface density, we have created a sequence of models from $n=0$ to $n=1.5$ which are normalized to reproduce the observed $7.0$~M$_J$ planet at $5.4$~AU.  As shown in Fig.~\ref{fig:sigma}, the latter implies that also the gas surface densities should be the same at $5.4$~AU. The model parameters are given in Table~\ref{Models}. We generally note that models with higher $n$ require a somewhat reduced value of $\alpha_L$, i.e. a reduced amount of specific angular momentum, as a larger fraction of the mass is in the interior. The required value of $\alpha_M$ decreases with decreasing $n$, since the normalization is based on the gas density at $5.4$~AU.  In the models with flat power-law slopes, the gas surface density is reduced on smaller scales, implying that a lower temperature needs to be reached for gravitational instability and fragmentation, while for steep power laws, fragmentation can occur at very high temperatures (Fig.~\ref{fig:temp}). The final clump masses are given in Fig.~\ref{fig:clumpf}, implying that the distribution becomes flatter if the gas surface density is steep, since it leads  to a larger gas reservoir in the interior. The ratio of the observed planetary masses implies $n\sim1$, though we cannot rule out the possibility that migration has occurred.
}

{We also note here that the final mass scale $M_f$ and the minimum mass for gap opening $M_{gap}$ are very similar, thus requiring further investigation of whether the planets can indeed clear their orbit. As shown in section~\ref{photoheat}, radiative feedback may, however, strongly contribute to stabilizing the disk on small scales. In the absence of gravitational instabilities, it is  much easier for the planets to clear their orbits and avoid migration into the central region \citep[e.g.][]{Baruteau11}. While the strength of radiative feedback in these systems is still uncertain, we  show below that there is a reasonable parameter space where heating from the central star will be important.}

{As an additional possibility, we  stress  that planets formed before the common envelope phase may also grow in these disks, with the final mass scale given as $M_f$. In fact, the final mass scale appears likely to be independent of the formation mechanism of the initial core, and whether it appeared owing to fragmentation or  to a pre-existing planet. The latter is still possible, since the analysis of \citet{Volschow13} and \citet{Mustill13} did not account for the potential mass and angular momentum accretion of the planets. In this sense, the pre-existing planets may be contained in the new planets as their central core.}

\section{Impact of stellar radiation}\label{rad}
Because the envelope of the AGB star is removed as a result of the ejection, the atmosphere of the remnant is likely to have a considerably higher temperature after the ejection event. Since the ejection radius is close to the transition towards the stellar core, the temperature of the stellar atmosphere $T_*$ can be substantially enhanced compared to the AGB stage. As the new stellar radius, we adopt the radius $r_{ej}$ at which the ejection of the envelope has occured. The stellar luminosity is then\begin{equation}
L_*=4\pi r_{ej}^2\sigma_{SB} T_*^4,\label{L*}
\end{equation}
where $\sigma_{SB}$ denotes the Stefan-Boltzmann constant. In this section, we explore the implications of the resulting stellar feedback on planet formation.

\subsection{Photoheating} 
\label{photoheat}
The radiation of the star may substantially contribute to the heating of the disk. Following \citet{Chiang97}, the interior disk temperature {in disk regions that are highly optically thick is given as} \begin{equation}
T=\left( \frac{\theta}{4} \right)^{1/4}\left( \frac{r_{ej}}{r} \right)^{1/2}T_*,
\end{equation}
where $T_*$ denotes the atmospheric temperature of the star, and $\theta$ is the grazing angle at which the light from the star strikes the disk. {The latter is given approximately  as $\theta\sim0.4 r_{ej}/r$.} {In regions where the photoheating is inefficient, we employ the critical temperature that is required} for the onset of fragmentation ($Q=1$) as derived in the previous section. 

{For our fiducial model C (power-law slope $n=1$ as defined in Table~\ref{Models}), we calculate the impact of different stellar surface temperatures $T_*$ on the gas temperatures of the disk as illustrated in Fig.~\ref{fig:temp_rad}. While for $T_*=10^5$~K, the heating is still significant even on scales of $4$~AU, the impact is more restricted towards $\sim1$~AU at $10^4$~K. In both cases, a significant part of the disk will be stabilized against fragmentation instabilities due to the effect of photoheating. The latter may provide a natural point to stop the inward migration, because planets can more easily evacuate their orbit from gas in the absence of turbulence and gravitational instabilities \citep{Baruteau11}. }

{We also note that, while these parts of the disk are stable against fragmentation, planets may still exist and grow in that region. The latter may occur either because of fragments migrating inward from larger scales or of pre-existing planets that have formed before the common envelope phase and have now increased their orbit owing to the mass loss of the binary. This, however, implies that the new planets may in fact contain the previous planets as a core. }

\begin{figure}[htbp]
\begin{center}
\includegraphics[scale=0.5]{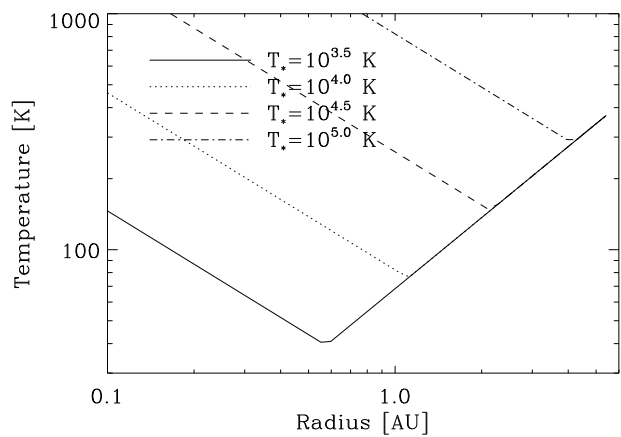}
\caption{{The gas temperature in the presence of photoheating from the star. We adopt here our fiducial disk model C as defined in Table~\ref{Models} and explore how the gas temperature depends on the temperature of the central star $T_*$. We further employ a minimum temperature defined from the condition of fragmentation instability ($Q=1$). }}
\label{fig:temp_rad}
\end{center}
\end{figure}

\begin{figure}[htbp]
\begin{center}
\includegraphics[scale=0.5]{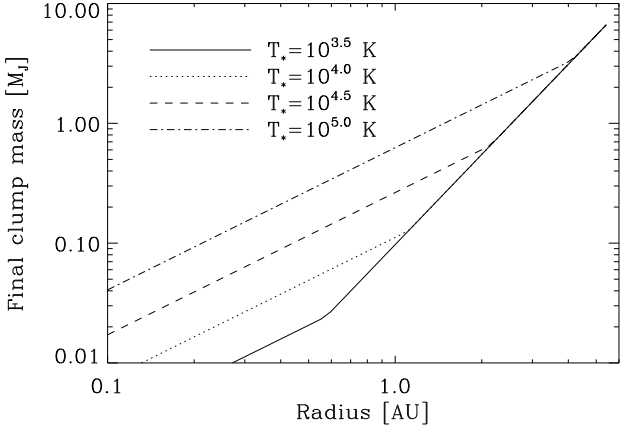}
\caption{{The final clump masses in the presence of photoheating from the star. We adopt here our fiducial disk model C as defined in Table~\ref{Models} and explore how the gas temperature depends on the temperature of the central star $T_*$. We further employ a minimum temperature defined from the condition of fragmentation instability ($Q=1$).
}}
\label{fig:temp_rad}
\end{center}
\end{figure}

\subsection{Photoionization}
As a result of the  high stellar temperatures, a substantial number of ionizing photons can be produced. Using Wien's law, we estimate the production rate of ionizing photons as
\begin{equation}
\dot{N}=\frac{L_*}{2.821 k_B T_*}.
\end{equation}
The size of the ionized region can be calculated by balancing the number of ionizations per second with the number of recombinations per second under the assumption of an isotropic distribution. We obtain \begin{equation}
\dot{N}=\int_0^{r_{HII}}4\pi r^2 dr n^2(r) \beta(T),
\end{equation}
where the recombination coefficient $\beta$ is given as\begin{equation}
\beta(T)=2.\times10^{-10}\ \mathrm{cm}^3\ \mathrm{s}^{-1} \left(\frac{T}{K} \right)^{-3/4}
\end{equation}
and the number density $n$ is estimated as $n=\rho/2m_p$, with $\rho=\Sigma/2h$. {For our fiducial model C, we} have  checked that a potential HII region remains negligible at least up to stellar temperatures of $10^{7.5}$~K and will not affect the formation of planets.

\subsection{Radiation pressure}
In addition to the photoheating and -ionization,  the momentum of the absorbed photons can also influence the evolution of the disk. The strongest impact can be expected in the optically thick limit, where the injected linear momentum per unit time is\begin{equation}
\frac{dp}{dt}=f_d\frac{L_*}{c},
\end{equation}
where $f_d=(2r\pi \cdot 2 h)/(4\pi r^2)=h/r$ is a geometrical factor that determines the fraction of photons that is aborbed in the disk plane. At the largest radii, we have $f_d=h/r=0.1$. The linear momentum obtained within a Kepler time $\tau_K$ then follows as\begin{equation}
p\sim f_d\tau_K\frac{L_*}{c},
\end{equation}
and can be cast as $p=M_{disk}v_r$, where $v_r$ denotes the radial velocity due to the injected momentum. Since the stellar luminosity depends strongly on the temperature of the atmosphere, the same is true for the resulting radiation pressure. At $T_*=10^{5.5}$~K, we find $v_r=f_d L\tau_K/(cM_{disk})\sim0.46$~km~s$^{-1}$. The latter is only marginally relevant and considerably lower than the escape velocity, which is given as $15$~km/s on a scale of $5$~AU. {The effect of photoheating means that the disk is  expected to be stable against fragmentation, while pre-existing cores or fragments migrating inwards from larger scales could still accrete matter. We note here that, while radiation pressure contributes no angular momentum, it may still alter the orbit of the gas, potentially favoring planet formation on larger scales.} With a typical growth rate of $\dot{M}_{cl}\sim M_{cl}\Omega_K$, the initial clumps can then be expected to grow to the observed values in only a few orbits. 

{Even for reduced values of $T_*$, we expect that radiation pressure could affect the surface layers of the disk and contribute to a redistribution of gas. In these layers, the interaction of dust, radiation, and gas could be of further importance \citep[e.g.,][]{Garaud04, Glassgold04}.}

 \begin{table*}[htdp]
\begin{center}
\begin{tabular}{c|c|c|c|c|c}
System & $M_{1,obs}$~$[M_\odot]$ & $M_2\ [M_\odot]$ & $M_P\ [M_J]$ & $r_P\ [AU]$ & $r_{P,cons}\ [AU]$\\ \hline
Detached sdB+MS/BD PCEBs\\ \hline
HW Vir & $0.485$ & $0.142$ & $6.8$ & $10.1$ & $1.0$\\
HS 0705+6700 & $0.483$ & $0.134$ & $6.7$ & $9.2$ & $0.94$\\
HS 2231+2441 & $0.47$ & $0.075$ & $5.3$ & $3.3$ & $0.34$\\
NSVS14256825 & $0.46$ & $0.21$ & $7.5$ & $21.9$ & $2.2$\\
NY Vir & $0.459$ & $0.122$ & $6.2$ & $8.5$ & $0.87$\\
\hline
Detached WD+MS PCEBs\\ \hline
NN Ser & $0.535$ & $0.111$ & $6.6$ & $5.4$ & $0.055$\\
V 471 Tau & $0.84$ & $0.93$ & $16.9$ & $90.9$ & $9.3$\\
QS Vir & $0.78$ & $0.43$ & $13.5$ & $30.1$ & $3.1$\\
RR Cae & $0.44$ & $0.183$ & $7.0$ & $18.7$ & $1.9$
\\ \hline
CVs\\ \hline
UZ For & $0.71$ & $0.14$ & $8.6$ & $4.9$ & $0.51$\\
HU Aqr & $0.80$ & $0.18$ & $6.3$ & $10.2$ & $0.64$\\
DP Leo$^1$ & $1.2$ & $0.14$ &$12$ & $1.8$ & $0.19$\\
DP Leo$^2$ & $0.6$ & $0.09$ & $6.6$ & $3.0$ & $0.30$
\end{tabular}
\end{center}
\caption{Predictions for planetary masses and orbits in the post-common envelope systems marked by \citet{Zorotovic13} as candidates for planets. We focus here on the most massive planet to form in a given system. $M_{1,obs}$ denotes the mass of the primary and $M_2$ the mass of the companion. $M_P$ is the expected mass of the planet. While $M_P$ is independent of the parameter $\alpha_L$, the orbit depends on its value. We therefore give the value $r_P$ for our fiducial scenario ($\alpha_L=12.5$) and a more conservative case $r_{P,cons}$ ($\alpha_L=4$). For the system DP Leo, the index $^1$ refers to stellar masses as given by \citet{Pandel02} and \citet{Beuermann11}, while $^2$ is based on the masses given by \citet{Schwope02}.}\label{tab}
\end{table*}%

\section{Application to other systems}\label{other}

Owing to the recent observations of \citet{Beuermann13} and the analysis by \citet{Volschow13} {and \citet{Mustill13}}, the model proposed here appears particularly relevant for NN Ser, because it seems  unlikely that the observed planetary orbits may result from a previous generation of planets {without accretion of additional mass and angular momentum}. Tto explore the relevance of this scenario for other systems, we  applied the model proposed here to the PCEB-systems marked by \citet{Zorotovic13} as candidates for planets, assuming that the interpretation of the LTT signals as caused by planetary companions is indeed correct. We note that, given the long orbital periods in combination with the rather short observational coverage, parameters of several of the planetary candidates are preliminary and may change. These candidates include five detached systems with a hot subdwarf B (sdB), four detached systems with a white dwarf (WD), and three cataclysmic variables (CVs). In all cases, the secondary corresponds to a main-sequence star (MS) or brown dwarf (BD). The data for the system NSVS14256825 currently allow for a one-planet solution on an excentric orbit \citep{Beuermann12b} or a two-planet solution on close-to-spherical orbits \citep{Zorotovic13}. The two-planet solution was, however, shown to be dynamically unstable \citep{Wittenmyer13}, while the one-planet solution was shown to be stable \citep{Beuermann12b}. For this system, we  thus deviate from \citet{Zorotovic13} by adopting the one-planet solution in the following.

Here, we employ the model parameters of $n=1$, $\alpha_M=1.1$, $\alpha_L=12.5$, and a mass loss factor $\mu=0.535/2\sim0.27$ as determined for NN Ser. This is a simplification to some extent, since the mass loss factor $\mu$ may vary for different systems. In addition, the orbital radius  depends on the parameter $\alpha_L$. We also consider a more conservative case with $\alpha_L=4$, leading to a more compact orbit while leaving the planetary mass unchanged. The results of these calculations are summarized in Table~\ref{tab}. For our initial comparison, we focus on the most massive planet in each system.

In the systems including a WD, a particularly good agreement with the masses given by \citet{Zorotovic13} is achieved for the systems NN Ser and RR Cae, where the difference in mass is less than a factor of 2 and the orbital radius in between the cases considered here. For V~471~Tau and QS~Vir, our model underpredicts the masses by a factor of 3-5. This could be  due to the uncertainties in our model, including the mass loss factor, but also hint towards a different formation mode in these systems.

\begin{figure}[htbp]
\begin{center}
\includegraphics[scale=0.5]{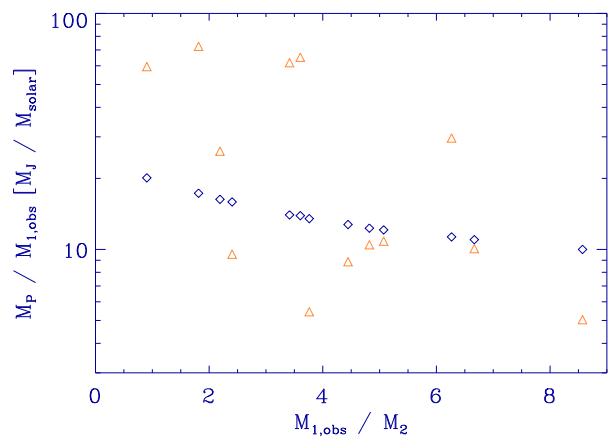}
\caption{The most massive planet in each system normalized in terms of the primary mass vs the ratio of primary to secondary mass. Diamonds show our model predictions and triangles observational data by \citet{Zorotovic13}. The data indicate the presence of a population consistent with our model predictions, but also five systems with systematically higher masses. The latter may correspond to a population that formed before the ejection event.}
\label{fig:stat}
\end{center}
\end{figure}

For the PCEBs, including a sdB, the planet in NSVS14256825 is close to the theoretical prediction, though slightly more massive by a factor of $1.5$. For NY~Vir, the planetary masses are a factor of $2$ below the derived values, which can be due to a lower mass loss factor or to additional fragmentation. In case of HW~Vir, HS~0705+6700 and HS~2231+2441, the observed masses are higher by a factor of 3-6, hinting again towards a higher mass loss factor or a different formation mode. For the CVs, we find very good agreement in all three cases, with mass differences of less than a factor of 2.  For the system DP Leo, we note that \citet{Schwope02} provide different stellar masses than \citet{Pandel02} and \citet{Beuermann11}, but the resulting planetary masses are both comparable to the  value derived by \citet{Zorotovic13}.

In the case of planet formation from the ejecta of common envelopes, the planetary masses can be expected to depend on the mass of the primary $M_1$ before ejection, given as $M_1=M_{1,obs}/\mu$ with $M_{1,obs}$ its current mass, and the mass of the secondary $M_2$. Assuming a constant mass loss factor $\mu$, we plot the mass of the planet normalized in terms of $M_{1,obs}$ as a function of the mass ratio $M_{1,obs}/M_2$ for both our theoretical predictions and the observed data given by \citet{Zorotovic13} in Fig.~\ref{fig:stat}. As stated before, we focus here on the most massive planet in each system. DP Leo is included using both the stellar masses given by \citet{Pandel02} and \citet{Beuermann11}, as well as by \citet{Schwope02}. We find that the theoretical predictions mark a narrow curve in the parameter space, and even 7 of our sources (8 data points due to DP Leo) are close to this curve. At the same time, the systems HW Vir, HS 0705+6700, V 471 Tau, and QS Vir show systematically higher planetary masses, with $M_P / M_{1,obs}>60 M_J/M_\odot$, and the system HS~2231+2441 has an enhanced value with $M_P / M_{1,obs}\sim30 M_J/M_\odot$.

{We emphasize  that the comparison pursued here is still prelimary. For instance, the common envelope ejection mechanism is not fully understood, and it is conceivable that model parameters can vary at least by a factor of 3. It is particularly important to develop models that consider deviations from spherical symmetry, the time evolution of the energy injection, the injection of angular momentum during the CE, and the angular momentum distribution prior to the CE. At this stage, it is thus too early to conclude whether a particular planet has been formed from the ejecta of the common envelope. However, at least in the case of NN Ser, it appears difficult to reconcile the current orbits with pre-existing planets if mass and angular momentum accretion is not considered \citep{Volschow13, Mustill13}. While the origin of the initial core most likely has no strong impact on the final mass scale, a more relevant uncertainty are the properties of the self-gravitating disk. In this respect, it is fairly encouraging that Fig.~\ref{fig:stat} hints at a population that is consistent with the model proposed here, while it also shows signs of a potentially different population of more massive planets. These possibilities should be addressed in more detail in future studies to both obtain  more solid knowledge of the properties of the planets and of the physical processes occurring during the common envelope phase and the subsequent evolution.
}

{We further investigate whether the properties of a second planet found in some systems can provide additional hints on the formation mechanism, even though we emphasize that migration cannot be ruled out at this point, and in fact some migration may be required to explain the 2:1 resonance reported by \citet{Beuermann13} for NN Ser.  In most of the two-planet systems}, the more massive planets are at larger radii (HW Vir, NY Vir, NN Ser, QS Vir), consistent with our model predictions for the final mass of the planets. An exception seems to be present for the system UZ For, where the interior planet is slightly more massive. The latter can possibly hint at additional fragmentation preventing the planets from reaching their final mass scale. We note again that we do not consider the two-planet solution for NSVS14256825, because it was shown to be unstable \citep{Wittenmyer13}.

For a more detailed comparison, we now calibrate the parameter $\alpha_L$ based on the largest observed planetary orbit, and calculate the expected planetary masses at the observed orbits. The expected and observed masses of the planets are given in Table~\ref{tab2}. 

\begin{table}[htdp]
\begin{center}
\begin{tabular}{c|c|c|c|c|c}
System &  $\alpha_L$ & $m_{th,1}$ &  $m_{o,1}$ & $m_{th,2}$ & $m_{o,2}$  \\ \hline
HW Vir & $12.5$ & $6.8$ & $30-120$ & $0.6$& $14.3$\\
NY Vir & $9.5$ & $6.2$ & $2.5$  & $2.4$ &  $2.3$\\
NN Ser & $12.5$ & $6.6$ & $7.0$ & $1.9$ & $1.7$\\
QS Vir & $6.0$ & $13.5$ & $56.59$ & $9.0$ & $9.01$\\
UZ For & $13.5$ & $8.6$ & $7.7$ & $1.4$ & $6.3$
\end{tabular}
\end{center}
\caption{Comparison of predicted and observed planetary masses in PCEBs with more than one planet. Here, $\alpha_L$ is calibrated based on the largest observed orbit, $m_{th,1}$ and $m_{th,2}$ are the expected planetary masses at the locations of the observed orbits, and $m_{o,1}$, $m_{o,2}$ are the observed masses of the planets.}\label{tab2}
\end{table}%

As noted before, the system NN Ser yields an excellent match between theoretical and observed values. In NY Vir, the most massive planet is smaller by about a factor of 2 than predicted, while the second planet is larger by a similar factor. Such fluctuations are rather typical for a statistical fragmentation process. For UZ For, the mass of the first planet matches very well, while the mass of the second planet is underestimated by about a factor of 5. While some fluctuations are indeed expected, this deviation is somewhat significant and could potentially hint at a planet from a previous generation. Larger deviations are seen for HW Vir and QS Vir, where the masses of the most massive planets are underestimated by about a factor of $5$. For HW Vir, also the second planet is considerably more massive than predicted, and we have checked that even another calibration of $\alpha_L$ according to the second planet leaves a discrepancy that is greater than a factor of $2$. For QS Vir, the mass of the second planet matches our theoretical expectations, therefore allowing for the possibility of a mixed population within one system. 

{Again, we  emphasize that the model predictions can vary at least by a factor of a few, and more detailed investigations of the scenario outlined here will be required in the future. The latter includes both the possibility of planet formation due to fragmentation and accretion, as well as the accretion and growth of pre-existing planets. We expect that these possibilities can hardly be distinguished by observational means, since the final mass scale is very likely close to the typical mass available on the orbit. A more detailed investigation employing hydrodynamical simulations may  clarify this point. It may, however, be possible to constrain the disk parameters from future studies, because these are strongly reflected in the resulting mass scale. 
 }

\section{Discussion and conclusions}\label{discussion}
In this study, we have explored the formation of planets from the ejecta of a CE via gravitational instabilities. For this purpose, we  adopted the model by \citet{Kashi11} to estimate the mass loss and the fraction of mass that remains gravitationally bound to the system. We further explored the properties of the resulting protoplanetary disk and estimated the characteristic mass scale of the planets. 

To produce the planets on orbits of $\sim5.4$~AU, as observed for NN Ser, we noticed that an {enhanced specific angular momentum in the disk is required compared to the specific angular momentum in the overall binary. The latter may  result from an inhomogeneous injection of the angular momentum during the inspiral of the companion star.} In addition, we note that the rotation of the CE may provide an additional source of angular momentum for the disk. {For systems like NN Ser, the model of \citet{Bear10} suggests rotation at up to $45\%$ of the breakup-velocity.} Under these conditions, the observed planetary masses can be reproduced, and our model predicts that more massive planets form at larger radii, as found in the observations by \citet{Beuermann10, Beuermann13}. 

{While some migration of the planets may be required in order to explain the 2:1 resonance reported by \citet{Beuermann13}, the effect of photoheating may contribute to stop planet migration on scales of a few AU, because  gap opening becomes more straightforward if gravitational instabilities are suppressed \citep{Baruteau11}. While the formation of new fragments would be suppressed in such a scenario, we note that pre-existing planets could still accrete the available gas reservoir and grow to a similar mass scale. Alternatively, the fragments may form on larger scales and open the gap in the regime where photoheating is efficient. 

From an observational point of view, it seems rather challenging to distinguish these possibilities. To investigate whether differences could occur in scenarios based on pre-existing planets or cores forming from new fragments, detailed hydrodynamical investigations will be required. A more important question may even concern the properties of the disk structure, and we expect that additional data on the current planet candidates may help constrain the properties of the underlying disks and their formation from the ejecta of the common envelope.} {As mentioned in the introduction, the young age of the white dwarf in NN Ser makes it unlikely that the planets have formed through the coagulation of planetesimals. We nevertheless mention that the pollution of white dwarfs from debris disks or asteroids has been reported in the literature \citep[e.g.,][]{Jura08, Jura09, Kulebi13}, and the growth of planetesimals may have occurred in older systems.

 Such processes would, however, need to compete with fragmentation via gravitational instability or the accretion of pre-existing planets, which occur on shorter timescales and may deplete part of the disk. Even around pulsars, rocky planets have, however, been found \citep[e.g.,][]{Wolszczan92, Konacki03, Yan13}, and the presence of a protoplanetary disk has been reported around the pulsar 4U\ 0142+61 \citep{Wang06, Ertan07}.}

With the model obtained for NN Ser, we  derived predictions for the planetary masses for an additional set of PCEBs identified by \citet{Zorotovic13} as candidates for planets. A comparison of our calculation with the observed masses hints at the presence of two populations, one that is consistent with our model predictions, and one with systematically higher masses, as in HW Vir, HS 0705+6700, V 471 Tau, and QS Vir. It is conceivable that these more massive planets have formed before the ejection event when a higher mass was available. Given the simplicity of our model, it is remarkable that a significant fraction of the population is reproduced without parameter adjustments and that  the masses of the second planet are also consistent in a large number of cases. 
We note that the planets in PCEB systems have been detected using the light-travel time (LLT) effect, which favors planets with high masses and large orbits. Our model predictions indeed show a characteristic trend towards high planetary masses and large radii, potentially explaining why the LLT method has been so successful. {In particular, we have shown that the highest final clump masses can be expected on scales of a few AU, while photoheating may in fact suppress fragmentation and gravitational instabilities in the interior, potentially preventing the migration of planets too close to the stars.}

For an improved understanding of PCEB planet formation, it is highly desirable to obtain additional data for a larger number of systems, but also to verify the orbits derived by \citet{Zorotovic13}  by measuring the lightcurves for additional years. In terms of the normalized planetary mass vs primary-to-secondary ratio diagram, we can then systematically assess whether  two populations are indeed present. In addition, a more sophisticated modeling is needed to account for potential differences in the mass loss factor, the fraction of retained mass and the distribution of the angular momentum. In this way, we can expect to improve our understanding of the planets in PCEB systems.

\begin{acknowledgements}
We thank {Robi Banerjee}, Klaus Beuermann, {Niels Clausen}, Rick Hessman, {Tim Lichtenberg, Farzana Meru}, Matthias Schreiber {and Andreas Tilgner} for stimulating discussions on the topic. DRGS and SD thank the {\em Deutsche Forschungsgemeinschaft} (DFG) for funding via the SFB~963/1 ``Astrophysical flow instabilities and turbulence'' (projects A5 and A12). DRGS further acknowledges financial support via the {\em Schwerpunktprogramm} SPP 1573 ``Physics of the Interstellar Medium'' under grant SCHL 1964/1-1.  {We thank the anonymous referee for valuable suggestions that helped {to improve} the manuscript.}
\end{acknowledgements}


\end{document}